\documentstyle[epsf,art12]{article}
\newcommand{\be}{\begin{eqnarray}}
\newcommand{\ee}{\end{eqnarray}}

\begin{document}
\baselineskip 0.320in
\begin{flushright}
\vskip -2cm
UAHEP974
\end{flushright}

\begin{center}
\vspace*{20mm}
{ \large \bf Scaling Violation through Squark and Light Gluino Production}
\\[7mm]
L. Clavelli\footnote{e-mail: lclavell@ua1vm.ua.edu} 
and I. Terekhov\footnote{e-mail: terekhov@lepton.ph.ua.edu} \\
{\it Department of Physics and Astronomy\\
The University of Alabama, Tuscaloosa Al 35487.}
\end{center}
\vskip 2cm
\begin{abstract}
\baselineskip 0.320in
In the light gluino scenario, squarks in the 100 GeV mass region can
be copiously produced at the Tevatron without a second heavy particle.
Their subsequent dijet decay into quark plus gluino leads to
non-scaling structure in the inclusive jet $X_T$ distribution. The
expected behavior is similar to recent observations.
\\11.30.Pb,14.80.Ly
\thispagestyle{empty}
\setcounter{page}{0}
\end{abstract}
\newpage
\par
 
Recent anomalies in the production of jets in $p \overline p$
annihilation have stimulated significant interest as a possible sign
of physics beyond the standard model. A case in point is the 
inclusive jet transverse energy cross section which was reported 
by CDF as exhibiting a dip at low $E_T$ and a rise at high $E_T$ 
relative to the standard model predictions.
\cite{ET}.  This behavior was cited as a possible indication of quark
substructure or of various other non-standard-model effects
\cite{CCS}. Among these latter was the suggestion that the jet $E_T$
distribution could be due, in the light gluino scenario, to extra jet
activity from production of gluino pairs and to the expected slower
running of $\alpha_s$ \cite{CT,BGM}. In addition to these two effects,
the possible production of a squark in association with a light gluino
could explain one of the several possible bumps visible in the CDF
data. (For a discussion of other indications of a light gluino see
\cite{Clavelli} and for a discussion of direct phenomenological
signals in future searches see \cite{Farrar}). On the other hand it
was also found possible \cite{CTEQ4} to fit the data, apart from the
low ($<50 $ GeV) $E_T$ values, by readjusting the gluon distribution
function in the proton in a way still consistent with other data 
or by changing the renormalization scheme \cite{KK}.
Thus, whether or not new physics is contained in the Fermilab data
must await further analysis. It is significant that
the angular distributions of the jets in various dijet mass bins are
consistent with that expected from the standard model
\cite{CDFangdis}. This would seem to rule out many non-standard-model
explanations of the data. However, it has been shown that the light
gluino hypothesis would lead to dijet angular distributions in
practice indistinguishable from the standard model expectations except
in dijet mass bins containing an up squark or down squark \cite{HRD}
\cite{Terekhov}. This is due to the fact that the structure of light
gluino production amplitudes is quite similar to that of other light
partons dominated by massless particle exchanges in the $t$ and $s$
channels. However a squark, once produced, will, in the light gluino
case, decay into dijets with an isotropic angular distribution in its
rest frame. 
\par 

The light gluino hypothesis, therefore, remains
viable only if the valence squarks are below about $150 $ GeV or above
$650 $ GeV since currently analyzed data does not constrain 
the lower or higher mass regions
\cite{TC,BGM,Terekhov}.
The possible bump at about $550 $ GeV seen by CDF and discussed in the
light gluino case in \cite{TC} might, therefore, be 
a statistical fluctuation. This interpretation is supported
by the failure of the later analysis \cite{Abachi,Abe} to confirm a
bump in this region. On the other hand the angular distributions have
not been published in the vicinity of a possible particle near $100 $
GeV suggested by the $E_T$ data. \cite{CT}. Furthermore $D0$ has not
published data spanning the relevant low $E_T$ region. Assuming it is
relatively less attractive to have squarks above $650$ GeV, the study
of jet angular distributions in the $100 $ GeV to $150 $ GeV dijet
mass region could therefore be crucial to the light gluino hypothesis.
\par

The parton distribution functions in the standard model must presently
be treated as theoretically arbitrary functions; the primary
constraints are from data on deep inelastic scattering and from direct
photon production. This freedom eliminates the necessity of, although
not the possibility of, new physics in explaining the Fermilab results
for the jet transverse energy cross sections at high $E_T$. It does
not at present, however, allow for an understanding of the behavior
below $50 GeV E_T$. In addition, it is possible to study suitably
defined scaling distributions whose ratio at two different Fermilab
energies is relatively insensitive to modifications of the parton
distributions. Such a quantity is the scaled inclusive jet transverse
energy cross section, which is predicted to have the form
\be
 s {d\sigma \over{dX_T}} = \alpha_s(\mu)^2 F(X_T,{\Lambda 
      \over{\sqrt{s}}},{m \over{\sqrt{s}}})
\label{sdsig}
\ee
with
\be
         X_T = {2 E_T \over{\sqrt{s}}}     .
\ee
Here $\mu$ is conventionally taken to be $E_T/2$, $\Lambda$ is the QCD
dimensional transmutation parameter, and $m$ represents any particle
mass appearing in the theory. Since the lowest order cross sections in
QCD are proportional to $\alpha_s^2$, this has been factored out,
although it also could be written as a function of the first two
arguments of the scaling function $F$. If all masses in the theory are
negligible compared to $\sqrt{s}$, $F$ depends on $s$ only through its
second argument coming from scaling deviations in the parton
distribution functions and from logarithmic scaling violations due to
higher order terms in $\alpha_s$ both of which effects are expected to
be  small. The ratio of the scaling distribution at two different
(high) energies is therefore expected to be approximately constant in
$X_T$ (for $X_T \stackrel{>}{\sim} 0.05$) 
independent of small modifications of the
parton distribution functions. The CDF collaboration \cite{CDFXT} has
presented preliminary results for the ratio
\be
  r(X_T) ={{{s d\sigma /{dX_T}}\quad (\sqrt{s}=630)}\over{
            s d\sigma /{dX_T}}\quad (\sqrt{s}=1800)}
\label{r}
\ee
In the standard model $r(X_T)$ departs from unity due to the running
of the coupling constant in (\ref{sdsig}) and due to logarithmic scaling
violations in the parton distribution functions. This leads to a
predicted value for $r$ near $1.8$ approximately independent of $X_T$.
The data show a systematic tendency to be below this prediction. In
the light gluino case $\alpha_s$ runs more slowly than in the standard
model leading to a value for $r$ near $1.6$ again approximately
constant if the squarks are too high in mass to be copiously produced.
\par

The current data for the ratio of the transverse energy distributions
at two energies are preliminary. Nevertheless, the reported structure
in $r(X_T)$, if not ultimately attributable to systematic errors, is a
tantalizing indication of the existence of strongly interacting
particles whose masses are not negligible compared to $630$ GeV and
which lead to a strong effect from the third argument in the $F$ of
(\ref{sdsig}).  Weakly interacting particles, such as the $W$ and 
$Z$ bosons,
do not have a sufficiently high production cross section compared to
QCD jet production to affect the $r$ parameter significantly.
Similarly the top quark or the prevalent supersymmetry hypothesis with
pair produced heavy squarks and gluinos have production cross sections
too low to be helpful in the current context. In addition, such
squarks are not expected to have prominent dijet decay modes.
\par

On the other hand, in the light gluino scenario, which can be obtained
in the context of the constrained supergravity related SUSY breaking
model by setting the universal gaugino mass $m_{1/2}$ to zero, a
single heavy squark can be produced in association with a light gluino
leading to greatly enhanced squark production cross sections as
discussed in \cite{TC,CT}. In the light gluino case, but not in the
standard SUSY picture, a squark will have a predominantly dijet decay
into quark plus gluino. Such a squark would produce a dip in the r
ratio (before smearing due to experimental resolution and
hadronization) at approximately $X_T=m(GeV)/1800$ followed by a peak
at $m(GeV)/630$. There is in fact some indication in the data for such
a low $X_T$ dip followed by a peak at roughly three times higher
$X_T$. In the current work we adopt the light gluino hypothesis and
consider as in \cite{TC,CT,Terekhov} the lowest order standard model
processes together with the effect of sparticle production processes
\be
G G \rightarrow {\tilde G}{\tilde G} \label{GG}\\
Q {\overline Q} \rightarrow {\tilde G}{\tilde G}\label{QQbar}\\
Q G \rightarrow Q {\tilde G}{\tilde G}\label{QGG}  .
\ee
Processes (\ref{GG}) and (\ref{QQbar}), while increasing the jet 
activity by some $6\%$ do not have a significant effect on the 
$r$ ratio. The possibility of a squark intermediate state in 
process (\ref{QGG}) leads however to structure in $r$ as discussed 
above. The effect is shown in Fig. 1 for a
squark mass of $135$ GeV. The structure shown in the $r$ parameter
theory as a function of $X_T$ is due to an intermediate squark in the
process of (\ref{QGG}). One could, as in \cite{BGM,HRD}, use the expected 
gluino
distribution functions in the proton to produce single squarks via the
quark-gluino fusion reaction $Q {\tilde G} \rightarrow {\tilde Q}$. If
the gluinos are dynamically generated from the gluons in the proton,
such a treatment is an approximation (for a forward going, narrow
width squark) to the full $2 \rightarrow 3$ process in (\ref{QGG}) 
above.
\par

In lowest order QCD, the squark width is predicted to be
\be
   \Gamma_{\tilde Q} = 2 M_{\tilde Q}\alpha_s/3  \label{width}
\ee
This width would be increased somewhat by electroweak decays of the
squark and by QCD corrections to the hadronic decays. In the
theoretical curve (solid line) in Fig. 1 we have 
assumed a width 30\% greater than (\ref{width}).  The width could
be increased further 
to simulate crudely the effect of experimental resolution and
hadronization smearing,
but we prefer to leave a more detailed consideration of such effects 
to a later paper. Since the peaks, if they exist, sit on a 
steeply falling background, the effect of resolution would be to move
the observed peak upward.
Given the preliminary nature of the data and
uncertainties in the actual amount of smearing present in the data we
would not consider any squark mass between $100$ GeV and $140$ GeV
as counter-indicated at present. The possible bump in the data near
$X_T \sim 0.28$ could be fit by a squark of mass $180$ GeV but such 
a mass might be ruled out by dijet angular distribution 
considerations \cite{HRD,Terekhov}.  
A consistent treatment of standard model processes and
corrections to processes (\ref{GG}) and (\ref{QQbar}) at order 
$\alpha_s^3$ would also
reduce the peaks somewhat by lifting the value of the non-resonant
backgrounds although, in the absence of peaks, such corrections would
not appreciably affect the $r$ ratio. Such refinements are left to a
later more detailed paper which will be justified if the observed
departure from standard model expectations survives a continued
experimental study of possible systematic errors. In particular we
suspect that the  low data point near $X_T = .07$ might be due to
a smearing effect in the $630 $ GeV data from regions of low $E_T$
where $\alpha_s$ becomes large and the parton distribution functions
may not be well behaved. 
\par 

    Our main conclusion at this point is
that the light gluino hypothesis together with valence squarks in the
$100$ GeV to $140$ GeV region are in qualitative agreement with
current experimental indications. A precise fit must await further
understanding of systematic uncertainties in both the experiment and
the theory. It seems unlikely that the current magnitude of the
observed structure could be fit in any model involving pair production
of two heavy particles with coupling strength $\alpha_s$ or less or 
in any model with additional gauge bosons in the electroweak sector.
\par

This work was supported in part by the Department of Energy under
grant $DE-FG02-96ER40967$.

\begin{figure}[p]
\hskip 5mm
\epsfxsize=5in \epsfysize=5in 
\epsfbox{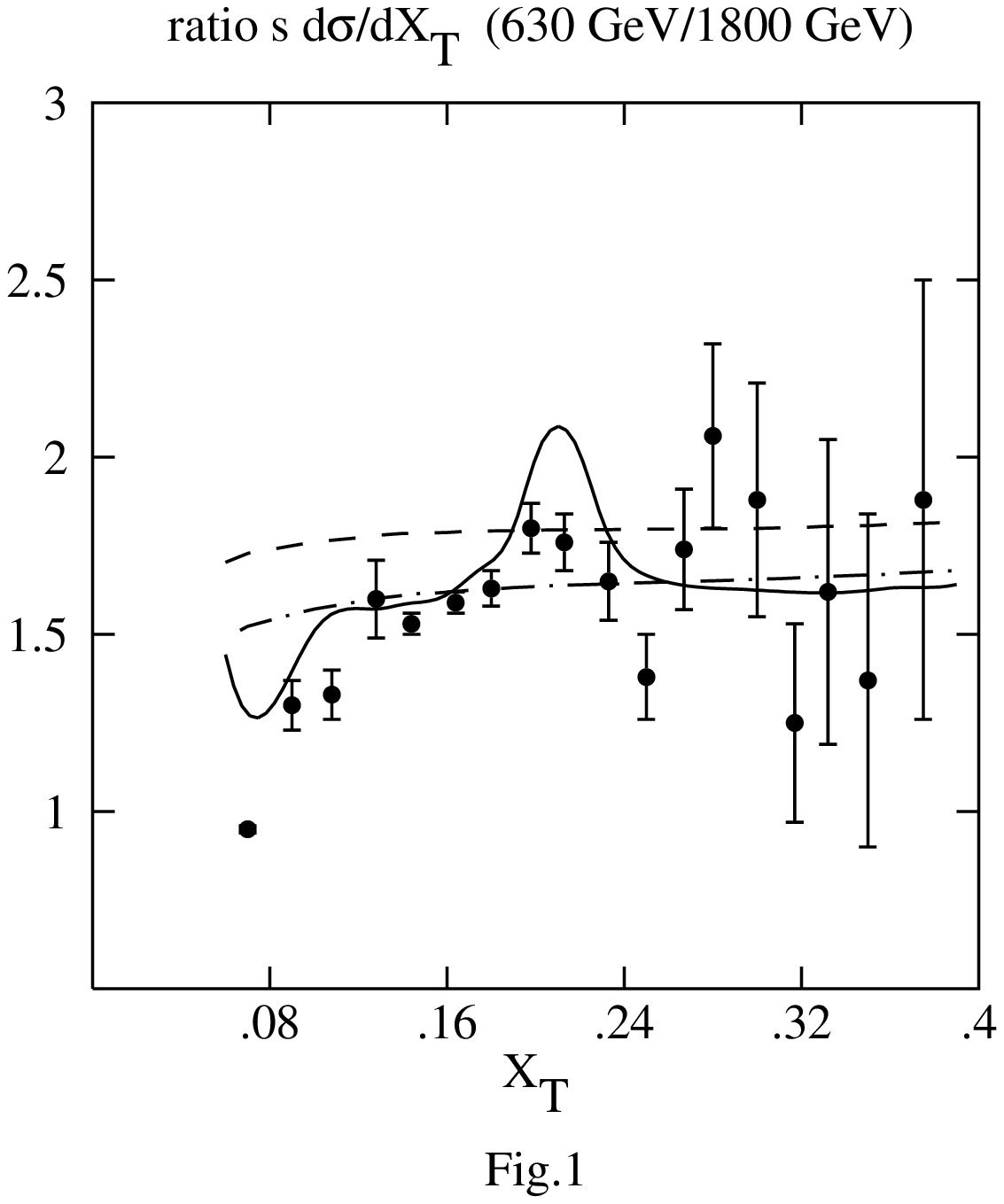}
\caption{
CDF data on the $X_T$ distribution compared to a) the standard model
prediction (dashed line), b) light gluino with decoupled (effectively
infinitely massive) squarks (dot-dashed line), and c) light gluino
plus $135 $ GeV valence squarks (solid line). 
}
\label{fig}
\end{figure}

\begin{thebibliography}{99}
\bibitem{ET} F. Abe et. al, CDF Collab., Phys. Rev. Lett. {\bf 77}, 
438 (1996).
\bibitem{CCS} R.S. Chivukula, A.G. Cohen, and E.H.
Simmons, Phys. Lett. {B380}, 92 (1996).
\bibitem{CT} L. Clavelli and I. Terekhov, Phys. Rev. Lett. {\bf 77}, 
1941 
(1996).
\bibitem{BGM} Z. Bern, A.K. Grant, and A.G. Morgan, Phys. Lett. 
{\bf B387}, 804 (1996).
\bibitem{Clavelli} L. Clavelli, Proceedings of the Workshop on the 
Physics of the Top Quark, IITAP, Iowa State Univ., Ames Ia, 1995, World
Scientific Press.
\bibitem{Farrar} G. Farrar, Phys. Rev. D51, 3904, 1995, Phys. Rev. Lett.
{\bf 76}, 4115 (1996).
\bibitem{CTEQ4} H.L. Lai, J. Huston, S. Kuhlmann, F. Olness, J. Owens, D.
Soper, W.K. Tung and H. Weerts, Phys. Rev. {\bf D55}, 1280 (1997).
\bibitem{KK} M. Klasen and G. Kramer, Phys. Lett. {\bf B386}, 384 (1996).
\bibitem{CDFangdis} F. Abe et al, CDF collab., Phys. Rev. Lett. {\bf 77} 
5336 (1996).
\bibitem{HRD} J. Hewett, T. Rizzo, and M. Doncheski, hep-ph/9612377 (1996).
\bibitem{Terekhov} I. Terekhov, hep-ph/9702301 (1997).
\bibitem{TC} I. Terekhov and L. Clavelli, Phys. Lett. 
{\bf B385}, 139 (1996).
\bibitem{Abachi} D0 Collaboration, S. Abachi et al., 
Phys. Rev. Lett. {\bf 75}, 618 (1995).
\bibitem{Abe} CDF Collaboration, F. Abe et al., Fermilab-PUB-97/023-E.
\bibitem{CDFXT} A. Bhatti, Fermilab-Conf-96/352-E, presented at the DPF
conference, Minneapolis, August (1996).
\end {thebibliography}
\end{document}